\documentstyle[11pt,newpasp,twoside]{article}
\def\edcomment#1{\iffalse\marginpar{\raggedright\sl#1\/}\else\relax\fi}
\reversemarginpar

\begin{document}
\title{Jets in Quasars}
\author{Marek Sikora}
\affil{N. Copernicus Astronomical Center, 00716 Warsaw, Bartycka 18, Poland}

\begin{abstract}
In my review of jet phenomena in quasars, I focus on
the following questions: How powerful are jets in radio-loud quasars? What is 
their composition? How are they launched? And why, in most quasars, are they 
so weak? I demonstrate the exceptional role that blazar studies can play in 
exploring the physics and structure of the innermost parts of quasar jets.

\end{abstract}

\section{Introduction}

The jet activity in quasars is  common, but very diverse.  
As radio observations indicate, jet powers
can differ by  several orders of the magnitude within the same
optical luminosity range. The most powerful jets
produce hundred kiloparsec-scale double radio structures. They can be
characterized as composed from
a pair of edge-brightened  radio lobes, with the  hotspots matching 
their luminosity peaks. Additionally, one sided jets are often observed,
 connecting one of two hotspots with the radio core in the center
of the host galaxy. The above  structure has a good  interpretation 
in terms of a dynamical model which involves propagation of light, 
relativistic jets in the IGM (Scheuer 1974; Begelman \& Coffi 1989). 
The hotspots are located at the ends of 
channels drilled by the jets through the IGM. They mark the regions
where material of the jet 
is shocked and spreads sideways, forming the radio lobe.
Relativistic speeds of jets explain their one-sided appearance,
while lightness (jet density lower than IGM density) is necessary to
explain formation of extended radio lobe structures
and the non-relativistic speeds of hotspots.

Highly polarized  and relatively steep radio spectra of extended radio
structures are uniquely interpreted in terms of optically thin synchrotron
radiation. The synchrotron spectra extend from  $\sim 10$ MHz, up to 
IR/optical. I discuss briefly, in \S 2, how these data can be  used to estimate the jet power.

Quasar jets can be  traced in radio down to parsec-scale central regions. 
There the jets are more relativistic 
(bulk Lorentz factor $\Gamma \sim 10$ 
[Padovani \& Urry 1992; Ghisellini et al. 1993; Homan et al. 2000]) than on 
kpc-scales ($\Gamma \sim 3$ [Wardle \& Aaron 1997]), 
and only those which are oriented close to the line of sight are bright enough
to be observed in detail.  Parsec-scale jets viewed ``pole-on''  can be 
decomposed into radio-cores and one-sided linear structure. The linear structure is very inhomogeneous, with
some bright regions propagating with relativistic  speeds and appearing to us as ``superluminal'' sources. Parsec-scale radio sources  show spectra with 
a low-energy break due to synchrotron-self-absorption. 
The frequency of this break is larger
the closer to the center one measures, and superposition of spectra from all
radio components give the characteristic flat radio spectrum, with a
energy spectral index $\alpha < 0.5$.  Quasars with such radio spectra are
called FSRQ (flat-spectrum-radio-quasars). I discuss the energetics and 
composition of quasar parsec-scale jets in \S 3.

As high frequency VLBI observations of nearby radio galaxies show, 
extragalactic jets are launched much deeper than the 
angular-resolution and synchrotron-self-absorption limited observations can
follow in quasars (Lobanov 1998; Junor, Biretta, \& Livio 1999). Fortunately, sub-parsec scale jets radiate a lot at higher, non-radio frequencies: up to 
optical/UV by the synchrotron mechanism, and in the X-ray and $\gamma$-ray bands via Comptonization of 
synchrotron  and  external diffuse radiation fields 
(Sikora, Begelman, \& Rees 1994; B{\l}a\.zejowski et al. 2000). 
This radiation, Doppler boosted 
into our direction, often dominates  over thermal quasar components, such as:
UV/optical 
radiation of the accretion disk, X-ray radiation
of a disk-corona, and IR radiation of dust located in a molecular torus
and heated by a disk. FSRQ which have spectra dominated by the 
non-thermal radiation from a jet are called blazars. Historically,
this category also includes  BL Lac objects, those sources with thermal 
signatures too weak present at all to place them in the quasar category. 
Hence, in order to avoid confusion while  talking about ``quasar-hosted'' 
blazars, I will call them ``Q-blazars''.

The broad-band spectra of Q-blazars can in general be superposed from 
the radiation produced  over a large distance range. However, 
at least during short-term high amplitude flares the spectra are dominated 
by radiation produced co-spatially, very likely in short-lived shocks,
somewehere at  0.1--1.0 pc  from the center.   Thus, as discussed in \S 4, multiwavelength studies of flares in blazars provide exceptional tools for 
exploring the structure and physics of jets on sub-parsec scales. 

Jets are predicted to produce radiation not only by relativistic electrons 
(here and after the term `electrons' is used for both electrons and positrons),
but also by cold electrons. Streaming with a bulk Lorentz factor 
$\Gamma \sim 10$, the cold electrons Compton scatter the external 
optical/UV photons and boost them up to 
 the soft X-ray range (Begelman \& Sikora 1987; Sikora et al. 1997).
It should be emphasized that cold electrons are an  unavoidable constituent
of jets near their base, where even mildly relativistic electrons cool 
faster than they propagate. And they can be present up to distances 
where non-thermal flares are produced, dragged by
as-yet-unshocked portions of the flow. 
As yet, no soft X-ray excesses have been 
confirmed. The  upper limits imposed on the number of cold electrons 
by observed soft X-ray fluxes exclude pure e$^{+}$e$^{-}$-jets and provide strong constraints
on the minimum distance of jet acceleration and collimation (Sikora \& Madejski 2000). 
These constraints, together with possible jet production scenarios, are discussed in \S 5.

Of course, any model of jet production should be able to explain
the huge range of jet powers.  Recent discoveries that many luminous radio-quiet quasars  reside --- like the radio-loud quasars --- in giant ellipticals (Taylor et al. 1996;
Kukula et al. 2000), and that the   
galaxy environments of the same luminosity radio-quiet and radio-loud quasars
is similar (McLure \& Dunlop 2000), challenge the previous claims 
that radio-loudness can be related to the morphology of the host galaxy or its clustering richness.  Furthermore, optical/UV spectral  similarities (Francis, Hooper, \& Impey 1993;
Zheng et al. 1997) and recent discoveries that BAL (broad absorption line) systems
exist also in radio-loud quasars (Brotherton et al. 1998; Becker et al. 2000)
suggest that radio-loudness is not very dependent on the parsec-scale environment, as well. 
All the above strongly supports the so-called spin paradigm, according to which powerful jets, giving rise to radio-loud quasars, can be produced  only with the help of rapidly rotating black holes.
How  the spin paradigm relates to central engine models
and the evolution of quasars is discussed toward the end of \S 5.

\section{Radio Lobes}

What can we  learn about jets from radio lobes? 
First of all, they provide very useful information about the energetics of 
jets,
which, contrary to that derived from radio-core scales and smaller, is not 
biased by such uncertainties as jet bulk Lorentz factor, dissipation 
efficiency and variability. The procedure for deriving the jet power is simple,
 but not free of assumptions and approximations. The first step is to recover from the observed electromagnetic spectrum the energy distribution of electrons.This can be done using the following approximate formula:
$$ L_{\nu,syn}d\nu  \simeq (N_{\gamma} d\gamma) m_e c^2 
\vert \dot \gamma \vert \, \eqno (1)$$ 
where 
$$ \vert \dot \gamma \vert \simeq 
{4 \over 3} {c \sigma_T u_B \gamma^2 \over m_e c^2 } \, \eqno(2)$$ 
is the rate of electron synchrotron energy losses;
$u_B = B^2/8 \pi$ is the magnetic energy density;
and $\nu \propto \gamma^2 B$.  For power-law synchrotron spectrum
$L_{\nu} \propto \nu^{-\alpha}$, formula (1) gives 
$N_{\gamma} \propto \gamma^{-s}$, where $s = 2\alpha +1$.
The above procedure is not sufficient to determine 
the normalization of the electron energy distribution, however. Additionally, one needs to know the  intensity of  the magnetic field, which can be  
estimated by assuming energy equipartition between electrons and magnetic fields. With this assumption, the electron (and magnetic) energy  content of the quasar radio lobes is found to be in the range 
$10^{59}-10^{61}$ ergs which, when divided  by  spectrally or 
dynamically determined ages  of the radio lobes, $t_{lobe} \sim 3 \times 10^7$
years, gives jet powers $10^{45} - 10^{47}$ ergs s$^{-1}$ 
(Rawlings and Saunders 1991).

There are several reasons why the  above estimates
should be considered  as  lower limits.
First, the equipartition  condition corresponds almost exactly
with the minimum total  energy of electrons and magnetic fields.
Second, from the observed radio spectra one can deduce that most of
the energy carried by electrons is contained in the low energy part of their
distribution. Assuming that the electron distribution has a break at an energy corresponding with the lowest observable frequency $\sim 10$ MHz (limited by reflection of
radio waves in ionosphere), one finds  that 
$\gamma_{min} \sim 500 / \sqrt{B/10\mu{\rm G}}$. Since there is no  proof 
that radiation has 
an intrinsic cutoff at 10 MHz, the adopted value of $\gamma_{min}$ can be
greatly  overestimated and the total electron energy content  underestimated. {Third, Rawlings \& Saunders assumed no energy contribution from protons.

How much might the jet powers be  underestimated due to the above assumptions?
Detection of X-rays from hotspots of several nearby radio galaxies and their
interpretation in terms of the SSC (synchrotron-self-Compton) process allowed
one to derive the magnetic field and electron energy densities without 
assuming  equipartition (see Wilson, Young \& Shopbell 2000 and 
references therein). Departure from the equipartition condition have been found to be very small. However, recent observations of X-rays produced around 
quasar radio lobes show that the pressure of the external gas is several times larger
than the pressure in the radio lobes obtained assuming equipartition between
electrons and magnetic fields and no proton contribution 
(Hardcastle \& Worral 2000).  This inconsistency cannot be resolved, even assuming that
the electron distribution extends down to $\gamma_{min}\sim 1$, 
if the equipartition condition is kept. There must be  significant 
departure from equipartition between electrons and magnetic fields 
and/or  the lobe pressure is dominated by  protons.
There are some observations suggesting that, indeed, the equipartition 
conditions can be violated in radio lobes, with particle
pressure dominant over magnetic pressure  (see Blundell and Rawlings 2000 and
references therein).  Yet another possibility is that extra
pressure in radio lobes is provided by cosmic rays accelerated via the Fermi process operating in the boundary layer between a jet and the surrounding medium (Ostrowski 2000). 

The rate at which  energy is delivered to radio lobes 
can be estimated more directly, just from the  bolometric 
luminosities of hotspots.  The rate is
$$ L_j = {L_{HS} \over \eta_e \eta_{rad}}  \, \eqno(3) $$
where $\eta_e$ is the fraction of kinetic energy of a jet converted
in the shock to relativistic electrons and $\eta_{rad}$ is the fraction
of electron energy lost by radiation. If, in the radiative regime, the
electromagnetic spectrum has a slope $\alpha \simeq 1$, 
then $\eta_{rad} \sim \ln(\nu_{max}/\nu_c)/\ln(\nu_{max}/\nu_{min})$,
where $\nu_c$ is the ``cooling'' break,  and $\nu_{min} \le 10$ MHz.
Applying this  for hotspot  D in Cyg A, where  
$\nu_c \sim 10^{10}$ Hz and $\nu_{max} \sim 10^{12}$ Hz, one can  find that
for $\nu_{min} = 10$ MHz, $\eta_{rad} \sim 0.4$. Now, assuming that
  the energy dissipated in the hotspots is equally shared by electrons and
magnetic fields, i.e., $\eta_e = 1/2$, and adopting from
Meisenheimer et al. (1997) $L_{HS} \simeq 4 \times
10^{44}$ ergs s$^{-1}$,
 we obtain $L_j \sim 2 \times 10^{45}$ ergs s$^{-1}$.
This estimate  of $L_j$ is consistent with that
deduced from the radio-lobe energetics (Carilli \& Barthel 1996). In both 
cases the energy is underestimated only by a factor $3/2$, if there are protons and they 
equally share energy with electrons and magnetic fields.  
Unfortunately, hotspot spectra up to the highest synchrotron
frequencies are currently available only for nearby radio galaxies and, 
therefore,
the bolometric-luminosity method cannot be applied to distant quasars.

What about the  pair content of  radio-lobe plasmas?
Noting that jets are approaching the hotspots with relativistic speeds,
the average energy of shocked protons is expected to be of the order
of the jet Lorentz factor. Thus, if the dissipated kinetic energy of a jet  
is initially shared equally  by electrons, protons, and magnetic fields, the average electron energy  should be 
$\gamma \sim (1/3) \Gamma (m_p/m_e) (n_p/n_e)$ $\sim 600\Gamma(n_p/n_e)$.
Since radio observations cannot follow electrons with energies lower
than $\gamma \sim 500 / \sqrt {(B/10 \mu{\rm G})}$, the pair 
content cannot be verified by radio lobe observations.

\section{Parsec-scale Jets}
In order  to recover the  physical parameters of radiating plasma in
 compact radio sources, one needs: to use the emissivity formula 
(like that in equation [1], but written in the source comoving frame);
to transform the comoving luminosity and frequency to the observed ones;
and to take advantage from two of the following:

a) the value of the bulk Lorentz factor $\Gamma$ (if available from
VLBI observations); 

b) X-ray flux, provided the X-rays are produced by the SSC process;

c) synchrotron-self-absorption break;

d) equipartition condition.

\noindent
Such analyses have been performed for a large sample
of compact radio sources by Ghisellini et al. (1992), who used (b) and (c),
and for the series of compact radio components in 3C 345 and 3C 279 
by Hirotani et al. (1999; 2000), who used (c) and (d).
Among other aspects, they calculated   electron energy distributions, 
$n_{\gamma} = C \gamma^{-s}$, assuming
$\gamma_{min}=1$, where $s = 2\alpha +1$. This allows one to obtain
electron energy densities, 
$u_e' \equiv  n_{\gamma}' \langle \gamma \rangle m_e c^2$, and then
energy fluxes of relativistic electrons 
$$ L_e \sim u_e' \Gamma^2  \pi a^2 c \,  \eqno (4) $$
where $a$ is the cross-sectional radius of the  source.
For the studied sources, values of $L_e$ are in the range 
$10^{45} - 10^{47}$ ergs s$^{-1}$, provided $\gamma_{min} = 1$, 
and 2-3 times smaller, if $\gamma_{min} \gg 1$.
 
Hence, the energy fluxes of relativistic electrons alone come close 
to satisfying the } energy requirements of radio lobes. However, one should
note that the total energy flux also includes  other forms of energy and,
in general, we have:
$$L_j = {L_e\over  \eta_{diss} \eta_e (1-\eta_{rad}) } \, \eqno (5)$$
where $\eta_{diss}$ is the fraction of the total jet energy which is
dissipated and used to accelerate electrons, to heat protons and to amplify
magnetic fields; $\eta_e$ is the fraction of the dissipated energy which
is used to accelerate electrons; and $\eta_{rad}$ is the fraction of 
electron energy lost by radiation during the source lifetime.
Thus, with the derived values of $L_e$, the total energy fluxes, $L_j$,
become dangerously high, particularly if  $\eta_{diss} < 0.1$ as intrinsic shock theories predict. Therefore, provided that the derived densities of relativistic electrons
are not affected by systematic errors   
(noting their very  strong dependence on 
the absorption-turnover frequency and the source geometry), 
one needs to postulate external  shock models,
with dissipation efficiencies $\eta_{diss}\ge 0.5$ (Dermer \& Chiang 1998). 

What is the pair content of the compact radio sources? 
Assuming that the dominant energy carriers are protons, we have
$$ L_j \simeq L_{p,0} + L_{diss} = L_{p,0} +  \delta L_p + L_B + L_e \, ,
\eqno(6) $$
where $L_{p,0} = n_p' m_p c^3 \pi a^2 \Gamma^2$ and $\delta L_p \simeq 
n_p' m_p (\langle \gamma_p \rangle -1)c^3 \Gamma^2 \pi a^2$.
Since
$${L_e \over L_{p,0}} = {n_e' \langle \gamma \rangle m_e \over n_p' m_p} =
{\eta_e \eta_{diss} (1-\eta_{rad}) \over 1-\eta_{diss}} \, \eqno (7) $$
where
$\eta_e = 1/(1 + L_B/L_e +\delta L_p/L_e)$, we obtain  
$$ {n_{+}' \over n_p'} \simeq {n_e \over 2 n_p} \simeq 
{50 \over \gamma_{min}} (3\eta_e){ \eta_{diss} \over 1-\eta_{diss}} 
(1-\eta_{rad}) \, \eqno(8) $$
where I used  $\langle \gamma \rangle \sim 6 \gamma_{min}$, which 
corresponds with $\alpha \simeq 0.6$.

Unfortunately, due to the synchrotron-self-absorption it is impossible to 
follow electrons with energies 
$\gamma < 50 \sqrt{ (\nu_{abs}/1{\rm GHz})/(B/0.1 {\rm G})}$ and determine
$\gamma_{min}$ directly from the observed spectra.
One can eventually try to determine the upper limits for the minimum electron 
energies from measurements of circular 
polarization provided such polarization results from the Faraday conversion mechanism. 
Measurements of circular polarization have been completed 
for several extragalactic sources (Homan \& Wardle 1999) and at least in 3C 279 there are 
indications that circular polarization is  produced by this mechanism
and that $\gamma_{min} < 20$ (Wardle et al. 1998). For such $\gamma_{min}$'s  
the number of pairs per proton can range from a few up to tens. 
There are several indirect arguments in favor of rather low pair contents:
(i) if pairs are produced in the central engine or its vicinity, then
their  flux is very limited by the annihilation process 
(Ghisellini et al. 1992);
(ii) if pairs are created  by nonthermal pair cascades operating in
the jet shocks, they would produce much softer X-ray spectra
than observed in Q-blazars (Ghisellini \& Madau 1996); (iii) jets with 
a large number of cold 
electrons would produce soft X-ray bumps by Comptonization of
UV disk radiation and BELs and such bumps have not been confirmed 
(Begelman \& Sikora 1987; Sikora et al. 1997).


\section{Q-blazars}
Strong and fast variability  in blazars is commonly interpreted in terms
of the shock-in-jet model. Producing a flare of the observed time scale
$t_{fl}$, the shock passes a distance range 
$$ \Delta r_{fl} \sim c t_{fl} \Gamma^2  \sim 100 
(t_{fl}/1 {\rm day}) (\Gamma/10)^2 \,\, {\rm lt-days} .   \, \eqno(9) $$
Sharp profiles of flares and comparable time scales of their rise and 
decay suggest that  shocks are launched at distances 
$r_{fl} \sim \Delta r_{fl}$ (Sikora et al. 2000). 
The distance of flare production can be also 
estimated  from the spectral location of the $\gamma$-ray luminosity peak,
provided the peak  is related to the break in the electron 
energy distribution caused by the cooling effect, i.e. that above the peak
electrons radiate on time scales shorter than the lifetime of the shock.
If $\gamma$-ray production is dominated by  the ERC (external-radiation-Compton)
process, and the luminosity of the ERC component is larger than the luminosity
of the synchrotron component, then the spectral distance is
$$ r_{sp} \sim c t_{sp}' \Gamma \sim
{m_e c^2 \over \sigma_T} \sqrt{\nu_{diff} \over \nu_c} 
{1 \over u_{diff}} \,  \eqno (10)$$ 
where the following relations were used: 
$t_{sp}' = \vert \gamma_c /\dot \gamma_c \vert$; 
$\vert \dot \gamma \vert \sim \Gamma^2 \gamma_c^2 \sigma_T u_{diff}/m_ec$;
and $\gamma_c \simeq \sqrt {\nu_c/\nu_{diff}}/\Gamma$.
It can be checked that both the variability distance, $r_{fl}$, and the 
spectral distance, $r_{sp}$,
are of the same order if $u_{diff} \sim 0.005$ ergs cm$^{-3}$, 
i.e., if $L_{BEL}(r_{fl}) \sim 10^{44}$ ergs s$^{-1}$,
and/or covering of the central source by dust at $T=1000$ K is $\sim 0.1$. 
Both
quantities are consistent with ourknowledge about BELR and near-IR radiation in quasars 
(B\l a\.zejowski  et al. 2000).

Equations (9) and (10), combined 
with  emissivity formulae for the  production of X-rays via the ERC
process (B{\l }a\.zejowski et al. 2000)  and with given value of 
$L_{diff}(r_{fl})$,
can be used to calculate the number of relativistic electrons involved
in flare  production and their energy flux, $L_e$.
For $L_{BEL} \sim 10^{45}$ ergs s$^{-1}$ and $\Gamma$ calculated
from the model, it can be  found that  $L_e \sim 10^{45} L_{SX,46}\,  {\rm ergs \, s}^{-1}$,
where $L_{SX}$ 
is the soft X-ray luminosity (Sikora et al., in preparation). 
Since  soft X-rays are very likely dominated 
by the SSC process (Inoue \& Takahara 1996; Kubo et al. 1998; B{\l}a\.zejowski
et al. 2000), the above estimate should be considered only
as the upper limit.

The electron energy flux, $L_e$, can be estimated also using  
the bolometric luminosity procedure. We have
$$ L_e \sim  {\Omega \over 4 \pi}
 {L_{QB} \over \eta_{rad}}   \, \eqno(11) $$
where $\Omega \sim \pi/\Gamma^2$ and $L_{QB}$ is the total apparent luminosity
of a blazar.
During high states $L_{QB}$ is dominated by luminosity in $\gamma$-ray bands
and is of the order $ 10^{48-49}$ ergs s$^{-1}$ (von Montigny
et al. 1995). From typical 
high energy spectra of Q-blazars ($\alpha \sim 1$ for $h\nu > 30 MeV$ 
and location of $\nu_c$ in the $1-30$ MeV range),
one can conclude that $\eta_{rad} \sim 1/2$ if $\gamma_{min} \sim 1$, 
and $\eta_{rad} \sim 1$ if  $\gamma_{min} \sim 100$. With these numbers, 
and $\Gamma$ taken from
the model calculations, we obtain 
$L_e \sim 1-2 \times 10^{45} (\Gamma/15)^{-2} L_{QB,48}$ ergs s$^{-1}$,
which is of the same order as calculated using the X-ray flux.
Noting that in Q-blazars
$L_e$ is about 1-2 orders lower  than $L_j$ estimated from radio lobes,
and that $L_e = \eta_{diss} \eta_e L_j (1-\eta_{rad})$, one can conclude that 
activity of the sub-parsec jets is governed by low dissipation efficiencies,
very likely by intrinsic shocks. This conclusion can actually be reinforced by 
the fact that the high (flaring and radio active) states are not permanent,
and after averaging over longer periods of time the values of $L_e$ are
likely to be lower by at least a factor two.
 
If the main carriers  of energy 
are cold protons, then from equation (8) the pair content is 
$ n_{+}'/ n_p' \sim 5 / \gamma_{min}$ for
$\gamma_{min} < 5$, and is  negligible for $\gamma_{min} > 5$, assuming
$\alpha = 0.6$, $\eta_{diss} = 0.05$ and $\eta = 1/3$.
Here, like in the radio lobes and compact radio sources, we have a sort of
conspiracy regarding the value of $\gamma_{min}$.
Since X-rays  from  blazars are presumably dominated by the SSC process, 
it is very difficult to follow the low energy portions of the ERC spectra 
to check whether there are any signatures of the low-energy 
cutoff in energy distribution of  electrons.
But at least  we can say that a large pair content $n_{+}/n_p \gg few$
can be excluded. One could speculate that there is a large
number of cold pairs.  However, this is excluded
because  such pairs, Comptonizing external UV photons,  would produce
a huge  soft X-ray bump, which is not confirmed observationally. 

\smallskip

\section{Central Engine and Spin Paradigm}

The absence of soft X-ray bumps in Q-blazar spectra   provides  
strong constraints on the jet's  structure near its base. There,
Comptonization of the disk  radiation by cold electrons, dragged
by the  relativistic and well collimated jet, 
is predicted to  produce a  prominent X-ray bump even if number of pairs
is zero. This suggests that powerful jets are  wider and/or  slower
 at their bases (Sikora \& Madejski 2000). They can be  launched by the 
innermosts parts of
the accretion disk, with the matter pulled from the disk surface 
and accelerated along the open magnetic field lines by centrifugal forces 
(Blandford \& Payne 1982). Such proto-jets are probably  collimated 
further away, by a disk corona or winds predicted by some models to be formed 
at distances $\ge 100 R_g$ (Rozanska \& Czerny 2000; Murray \& Chiang 1997;
Proga, Stone \& Kallman 2000). Yet in the acceleration zone, the initially proton-electron outflows can be loaded by e$^{-}$-e$^{+}$ pairs. This is
due to boosting of  the coronal hard X-rays by cold electrons in the outflow  
up to MeV energies and the subsequent absorption of MeV photons in the 
$\gamma\gamma$ pair production process.
Depending on the geometry and kinematics of the proto-jet, the number of 
pairs per proton can reach a value ranging from a few up to a few tens.

The fact that only 10 \% of quasars are radio-loud  implies that  powerful 
jets are rare. It is very likely that  the leading parameter which decides 
about the jet power is the spin of the black hole, defined as $A= J/J_{max}$ where  $J_{max} = GM^2/c$
 (Wilson \& Colbert 1995; Moderski, Sikora \& Lasota 1998). 
Perhaps the  fast rotation of the  black hole is 
necessary to heat the accretion disk. Due to this heating the innermost parts 
of the disk can be inflated and this  can help to generate large scale magnetic fields,
which are required both to  accelerate MHD outflows and to
link a disk with the rotating black hole and/or the gas plunging 
into the ergosphere (Meier 2000; Krolik 2000). Furthermore, extra heating of 
surface layers of the disk by a rotating black hole 
can help to put gas on magnetic field lines, if the latter  
are not bent enough to allow  centrifugal forces to pull the gas directly
from the photosphere.

Now, provided that the above scenario(s) gives a strong enough dependence of 
the jet  power on the spin of the black hole to explain the huge range of
radio-loudness,  the next question which should be answered is how  
Nature managed  to have only a small fraction of rapidly rotating black holes.
Let us recall that accretion disks, acting enough long to  
double the black hole mass, spin up the holes  to $A \sim 1$ 
(Bardeen 1970;
Thorne 1974). On the other hand,  once the black hole is spun up, it 
cannot be easily slowed down, certainly not during
the low accretion phases during which the rate of extraction of black 
hole energy is very small because  of its proportionality  to 
$B^2 \propto \dot M$. 
Hence, if the spin paradigm is right and the growth of supermassive black holes
is governed by the large accretion events, then
the number of  radio loud quasars should be larger than of radio quiet quasars,
oppositely to what is observed. A solution of  the problem can be that 
growth of the black hole in most objects is dominated by low-mass accretion 
events with random angular momentum orientations.
This picture is supported by observations
of Seyfert galaxies, which show that AGNs are randomly oriented 
relative to the host galaxy planes (Wilson \& Tsvetanov 1994;
Schmitt et al. 1996). Low-mass accretion events
are also supported by  analyses of the dynamics 
of Seyfert jet interactions with the host galactic gas which show that
the lifetime of these objects is of the order of $10^5$ years only 
(Capetti et al. 1999). 
Thus, it is tempting to speculate that only major mergers which involve at
least a one gas-rich galaxy can lead to the accretion disk operating long 
enough to double the black hole mass and
spin up the black hole to $A > 0.5$. During early phases of the process, 
when $A$ is still small, the quasar is predicted to show up as 
radio-quiet, and then slowly transform to a radio-loud one.
 After the fuel is off, the accretion drops,
but spin of the black hole remains high and such an object can 
eventually be represented by  FR I radio galaxies.

\acknowledgments
I am grateful to Mitch Begelman for his valuable comments which helped
improving the paper. This work has been supported in part by the Polish
KBN grant 2P03D 00415.


\begin{references}
\reference Bardeen, J.M. 1970, Nature, 226, 64 
\reference Becker, R.H., et al. 2000, ApJ, 538, 72
\reference Begelman, M.C., \& Cioffi, D.F. 1989, ApJ, 345, L21
\reference Begelman, M.C.,  \& Sikora, M. 1987, ApJ, 322, 650 
\reference Blandford, R.D., \& Payne, D.G. 1982, MNRAS, 199, 883
\reference Blundell, K.M., \& Rawlings, S. 2000, AJ, 119, 1111
\reference B{\l }a\.zejowski, M., Sikora, M., Moderski, R., Madejski, G.M. 
2000, ApJ, 545, 107
\reference Brotherton, M.S., et al. 1998, ApJ, 505, L7
\reference Capetti, A., Axon, D.J., Macchetto, F.D., Marconi, A. \& Winge, C.
1999, ApJ, 516, 187
\reference Carilli, C.L., \& Barthel, P.D. 1996, A\&AR, 7, 1
\reference Dermer, C.D., \& Chiang, J. 1998, NewA, 3, 157
\reference Francis, P.J., Hooper, E.J., \& Impey, C.D. 1993, AJ, 106, 417
\reference Ghisellini, G., Celotti, A., George, I.M., \& Fabian, A.C.
1992, MNRAS, 258, 776
\reference Ghisellini, G., \& Madau, P. 1996, MNRAS, 280, 67
\reference Ghisellini, G., Padovani, P., Celotti, A., \& Maraschi, L.
1993, ApJ, 407, 65
\reference Hardcastle, M.J., \& Worrall, D.M. 2000, MNRAS, 319, 562
\reference Hirotani, K., Iguchi, S., Kimura, M., \& Wajima, K. 1999,
PASJ, 51, 263
\reference Hirotani, K., Iguchi, S., Kimura, M., \& Wajima, K. 2000,
ApJ, 545, 100
\reference Homan, D.C., \& Wardle, J.F.C. 1999, AJ, 118, 1942
\reference Homan, D.C. et al. 2000, astro-ph/0009301
\reference Inoue, S., \& Takahara, F. 1996, ApJ, 463, 555
\reference Junor, W., Biretta, J.A., \& Livio, M. 1999, Nature, 401, 891
\reference Krolik, J. 2000, astro-ph/0008372
\reference Kubo, H., et al. 1998, ApJ, 504, 693
\reference Kukula, M.J., et al. 2000, astro-ph/0010007
\reference Lobanov, A.P. 1998, A\&A, 330, 79
\reference McLure, R.J., \& Dunlop, J.S. 2000, astro-ph/0007219 
\reference Meier, D.L. 2000, astro-ph/0010231
\reference Meisenheimer, K., Yates, M.G., \& R\"oser, H.-J. 1997, A\&A,
325, 57 
\reference Moderski, R., Sikora, M., \& Lasota, J.-P. 1998, MNRAS, 301, 142
\reference Murray, N., \& Chiang, J. 1997, ApJ, 474, 91
\reference Ostrowski, M. 2000, MNRAS, 312, 579
\reference Padovani, P., \& Urry, C.M. 1992, ApJ, 387, 449
\reference Proga, D., Stone, J.M., Kallman, T.R. 2000, ApJ, 543, 686
\reference Rawlings, S., \& Saunders, R. 1991, Nature, 349, 138
\reference R\'o\.za\'nska, A., \& Czerny, B. 2000, A\&A, 360, 1170
\reference Scheuer, P.A.G. 1974, MNRAS, 166, 513
\reference Schmitt, H.R., Kinney, A.L., Storchi-Bergmann, \& T., Antonucci, R.
1997, ApJ, 477, 623
\reference Sikora, M., Begelman, M.C., \& Rees, M.J. 1994, ApJ, 421, 153 
\reference Sikora, M., \& Madejski, G. 2000, ApJ, 534, 109
\reference Sikora, M., Madejski, G.,
Moderski, R., \& Poutanen, J. 1997, ApJ, 484, 108
\reference Taylor, G.L., Dunlop, J.S., Hughes, D.H., \& Robson, E.I. 1996,
MNRAS, 283, 930
\reference Thorne, K.S. 1974, ApJ, 191, 507
\reference von Montigny, C., et al. 1995, ApJ, 440, 525
\reference Wardle, J.F.C., \& Aaron, S.E. 1997, MNRAS, 286, 425
\reference Wardle, J.F.C., Homan, D.C., Ojha, R., \& Roberts, D. 1998, Nature,
395, 457
\reference Wilson, A.S., \& Colbert, E.J.M. 1995, ApJ, 438, 62
\reference Wilson, A.S., \& Tsvetanov, Z.I. 1994, AJ, 107, 1227
\reference Wilson, A.S., Young, A.J., \& Shopbell, P.L. 2000, ApJ, 544, L27
\reference Zheng, W., Kriss, G.A., Telfer, R.C., Grimes, J.P., \&
Davidsen, A.F. 1997, ApJ, 475, 496 
\end{references}
\end{document}